\documentclass[twocolumn,preprintnumbers,amsmath,amssymb,superscriptaddress]{revtex4-2}

\usepackage{mathtools}

\newcommand{\ket}[1]{|#1\rangle}
\newtheorem{proposition}{Proposition}

\begin{document}

\title{Efficient quantum compression for identically prepared states with arbitrary dimension}

 \author{Zeyu Chen}
 \email{chenzeyu@zju.edu.cn}
 \affiliation{School of Mathematics Science, Zhejiang University,
Hangzhou 310058, P. R. China}

\author{Chunhe Xiong}
 \email{xiongchunhe@zju.edu.cn}
 \affiliation{School of Mathematics and Statistics, Central South University, Changsha 410083, PR~China}

\author{Kamil Khadiev}
\email{kamilhadi@gmail.com}
\affiliation{Institute of Computational Mathematics and Information Technologies, Kazan Federal University, Kremlyovskaya, 35, Kazan, Russia}

 \author{Junde Wu}
 \email{corresponding author: wjd@zju.edu.cn}
 \affiliation{School of Mathematics Science, Zhejiang University,
Hangzhou 310058, P. R. China}

\begin{abstract}
Identical preparation creates permutation symmetry that can be used for lossless quantum compression. For $n$ copies of an unknown $d$-dimensional pure state, the tensor-power input lies in the fully symmetric subspace, whose dimension is polynomial in $n$ for fixed local dimension. Schur--Weyl duality isolates this subspace coherently, allowing the fixed representation labels to be discarded while retaining all information needed for recovery. The resulting encoder and decoder are exact on the input family. Because tensor-power states span the symmetric subspace, the achieved memory dimension is also necessary for reversible coherent compression, so the scheme is space-optimal. The encoder is implemented recursively with Clebsch--Gordan transforms, and only the branches reached by symmetric inputs need to be reproduced. Standard efficient synthesis of these transforms yields polynomial-size circuits for fixed local dimension and arbitrary target accuracy. Thus identically prepared pure states in arbitrary dimension admit lossless, space-optimal compression with an efficient circuit implementation.
\end{abstract}

\maketitle

\section{Introduction}

Quantum compression can exploit restrictions on the family of admissible input states. For $n$ identical copies of an unknown pure state $\ket{\phi}\in\mathbb{C}^d$, the joint state $\ket{\phi}^{\otimes n}$ is invariant under permutations and therefore lies in the symmetric subspace $\operatorname{Sym}^n(\mathbb{C}^d)$. This subspace is much smaller than the ambient tensor-product space: for fixed $d$ its dimension grows polynomially rather than exponentially with $n$. The compression problem is therefore to isolate these symmetric degrees of freedom by a coherent and reversible basis change, so that the original state can be recovered exactly.

Schur--Weyl duality gives the natural basis for this reduction. In the Schur decomposition, a tensor-power input occupies only the fully symmetric $U(d)$ representation indexed by $[n]$, while its permutation representation is trivial. The encoder is therefore the restriction of the Schur transform to this sector, and the decoder is its inverse. Compression has been studied theoretically and experimentally for qubit ensembles \cite{plesch2010efficient,rozema2014quantum}, while efficient implementations of the Schur and Clebsch--Gordan transforms are known in arbitrary dimension \cite{bacon2007quantum}. Together, these ingredients give the compression in arbitrary dimension, a matching memory lower bound, and an explicit implementation.

The symmetric restriction also identifies the part of the recursive Clebsch--Gordan transform needed by the input family. At the $k$th step, the input can only enter the $[k+1]$ output branch. On this branch, the Clebsch--Gordan coefficients have a simple occupation-number form, and the reduced-Wigner recursion collapses at each level to a controlled two-level rotation. This structure yields the restricted transform. Standard synthesis of the recursion then gives polynomial complexity for fixed $d$, while the mathematical compression map remains exact.

\section{The Schur transform and related representation theory background}

The Schur transform is the change of basis associated with the decomposition of the $n$-qudit space $(\mathbb{C}^d)^{\otimes n}$. The relevant representation-theoretic structure comes from the joint actions of $U(d)$ and $S_n$ on this space.

A representation of a group $\mathcal{G}$ on a vector space $V$ is a homomorphism $\pi:\mathcal{G}\to\mathrm{GL}(V)$ \cite{fulton2013representation,ping2002group}. Throughout, all representations are complex and finite-dimensional, and we often identify $(\pi,V)$ with its representation space $V$. For vector spaces $V_1$ and $V_2$, let $\mathrm{Hom}(V_1,V_2)$ denote the space of linear maps from $V_1$ to $V_2$. If they carry representations $\pi_1$ and $\pi_2$ of $\mathcal{G}$, respectively, the induced action $\pi$ on $\mathrm{Hom}(V_1,V_2)$ is
\begin{align}
    \pi(g)(\varphi) = \pi_2(g) \circ \varphi \circ \pi_1(g)^{-1}
\end{align}
where $g\in \mathcal{G}$ and $\varphi\in\mathrm{Hom}(V_1,V_2)$. If $V_1$ and $V_2$ are isomorphic as $\mathbb{C}[\mathcal{G}]$-modules, the representations $(\pi_1,V_1)$ and $(\pi_2,V_2)$ are equivalent, and we write $V_1\stackrel{\mathcal{G}}{\cong}V_2$. A representation $(\pi,V)$ is irreducible if its only $\pi$-invariant subspaces are $\{0\}$ and $V$. Let $\Lambda$ be a complete set of inequivalent irreducible representations of $\mathcal{G}$. Any reducible representation $(\pi,V)$ then decomposes as
\begin{align}
\pi(g) &\cong \bigoplus_{\lambda\in\Lambda}r_\lambda(g)\otimes I_{n_\lambda}\\
    V&\stackrel{\mathcal{G}}{\cong} \bigoplus_{\lambda\in\Lambda} V_\lambda \otimes \mathbb{C}^{n_\lambda}
\end{align} where $(r_\lambda,V_\lambda)$ are irreducible representations and $n_\lambda$ is the multiplicity of $V_\lambda$ in $V$.

Consider the natural actions of $S_n$ and $U(d)$ on the space $(\mathbb{C}^d)^{\otimes n}$:

\begin{align}
\mathbf{P}(s)\left|i_1 i_2 \ldots i_n\right\rangle &=\left|i_{s^{-1}(1)} \ldots i_{s^{-1}(n)}\right\rangle \\
\mathbf{Q}(U)\left|i_1 \ldots i_n\right\rangle &=U^{\otimes n}\left|i_1 \ldots i_n\right\rangle
\end{align}
where $s\in S_n$ is a permutation of $n$ objects, $s(i)$ denotes the action of $s$ on the label $i$, and $U(d)$ denotes the group of $d\times d$ unitary matrices.

Schur--Weyl duality characterizes the decomposition of the tensor space under the joint action of $U(d)$ and $S_n$ \cite{weyl1946classical,goodman2000representations}. The representations $\mathbf{P}$ and $\mathbf{Q}$ commute, and $(\mathbb{C}^d)^{\otimes n}$ decomposes into tensor products of irreducible representations of $S_n$ and $U(d)$, each with multiplicity zero or one. These irreducible representations are indexed by partitions $[\lambda]=[\lambda_1,\lambda_2,\ldots,\lambda_d]$ satisfying $\lambda_1\geq\lambda_2\geq\cdots\geq\lambda_d\geq0$ and $\sum_{i=1}^d\lambda_i=n$:
\begin{align}
\mathbf{Q}(U)\mathbf{P}(s)&\stackrel{U(d) \times S_n}{\cong} \bigoplus_{[\lambda] \in \mathcal{I}_{d, n}}\mathbf{q}_{[\lambda]}^d(U)\otimes \mathbf{p}_{[\lambda]}(s)\\
\left(\mathbb{C}^d\right)^{\otimes n} &\stackrel{U(d) \times S_n}{\cong} \bigoplus_{[\lambda] \in \mathcal{I}_{d, n}} \mathcal{Q}_{[\lambda]}^d \otimes \mathcal{P}_{[\lambda]}
\end{align}
where $\mathcal{I}_{d,n}$ denotes the set of partitions of $n$ into $d$ nonnegative parts satisfying $\lambda_1\geq\lambda_2\geq\cdots\geq\lambda_d$. Equivalently, there is a basis $\ket{{[\lambda]}}\ket{q_{[\lambda]}}\ket{p_{[\lambda]}}$, called the Schur basis, in which both the $n$-fold tensor action of $U(d)$ and the action of $S_n$ are block diagonal:
\begin{align}
\mathbf{Q}(U)|[\lambda]\rangle\left|q_{[\lambda]}\right\rangle\left|p_{[\lambda]}\right\rangle&=|[\lambda]\rangle\left(\mathbf{q}_{[\lambda]}^d(U)\left|q_{[\lambda]}\right\rangle\right)\left|p_{[\lambda]}\right\rangle \\
\mathbf{P}(s)|[\lambda]\rangle\left|q_{[\lambda]}\right\rangle\left|p_{[\lambda]}\right\rangle&=|[\lambda]\rangle\left|q_{[\lambda]}\right\rangle\left(\mathbf{p}_{[\lambda]}(s)\left|p_{[\lambda]}\right\rangle\right)
\end{align}
The corresponding change of basis from the standard computational basis $\ket{i_1,i_2,\dots,i_n}$ to the Schur basis is the \textit{Schur transform}. It converts a local qudit-level description into a global symmetry-adapted representation.

\subsection{Construction of the irreducible representations $\mathcal{Q}_{[\lambda]}^d$ and $\mathcal{P}_{[\lambda]}$}

The irreducible representations $\mathcal{Q}_{[\lambda]}^d$ and $\mathcal{P}_{[\lambda]}$ admit subgroup-adapted bases constructed recursively from their branching rules.

A basis of $\mathcal{Q}_{[\lambda]}^d$ is called the \textit{Gelfand--Tsetlin basis} \cite{gelfand1950finite}, while that of $\mathcal{P}_{[\lambda]}$ is called the \textit{Young--Yamanouchi basis}.

The polynomial irreducible representations of $U(d)$ that occur in $(\mathbb{C}^d)^{\otimes n}$ are indexed by partitions $[\lambda]\equiv[\lambda_{i,d}]\equiv[\lambda_{1,d},\lambda_{2,d},\ldots,\lambda_{d,d}]$ of $n$ into $d$ nonnegative parts satisfying $\lambda_{i,d}\geq\lambda_{i+1,d}$. Their subgroup-adapted orthonormal bases follow from the \textit{Weyl branching law} \cite{weyl1946classical,goodman2000representations}. Under restriction from $U(d)$ to $U(d-1)$, $\mathcal{Q}_{[\lambda]}^d$ decomposes into distinct $U(d-1)$ irreducible representations whose labels satisfy the \textit{betweenness conditions}:
\begin{align}
\lambda_{i,d}\geq\lambda_{i,d-1}\geq\lambda_{i+1,d}
\end{align}
If two partitions $[\lambda]$ and $[\mu]$ satisfy this condition, we say that $[\mu]$ \textit{interlaces} $[\lambda]$ and write $[\mu]\preceq[\lambda]$. A basis vector of $\mathcal{Q}_{[\lambda]}^d$ is then determined uniquely by the array $\{\lambda_{ij}\}$ along the subgroup chain $U(1)\subset U(2)\subset\cdots\subset U(d)$, where the irreducible representations of $U(1)$ are one-dimensional.

A convenient notation for this subgroup-adapted basis is the \textit{Gelfand--Tsetlin pattern} $(\lambda)$, a triangular array of $\frac{d(d+1)}{2}$ nonnegative integers $(\lambda)=\{\lambda_{ij}\}$ arranged as

\begin{equation}
(\lambda)=
\begin{pmatrix}
\lambda_{1,d} & \lambda_{2,d} & \cdots & \lambda_{d,d}\\
& \lambda_{1,d-1} & \cdots & \lambda_{d-1,d-1}\\
& & \ddots & \vdots\\
& & & \lambda_{1,1}
\end{pmatrix}.
\end{equation}

The Gelfand--Tsetlin pattern $(\lambda)$ is also denoted by $\left(\begin{aligned}
[&\lambda]_d\\
(&\lambda)_{d-1}\\
\end{aligned}\right)$, where $[\lambda]_d$ is the partition denoting an irrep in $U(d)$.

Similarly, irreducible representations of $S_n$ are indexed by partitions $[\lambda]$ of $n$. In the Schur--Weyl decomposition above, only partitions with at most $d$ nonzero parts occur. Their subgroup-adapted bases follow the branching rule $S_n\downarrow S_{n-1}$: $[\lambda]$ decomposes into distinct representations $[\mu]$ obtained by removing one box, so that $\sum_i\mu_i=n-1$ and $\lambda_i-1\leq\mu_i\leq\lambda_i$. We write this relation as $[\mu]\in[\lambda]-\square$. A subgroup-adapted basis of $\mathcal{P}_{[\lambda]}$ is therefore specified by $n$ partitions $\{[\lambda]_i\}_{i=1}^{n}$ with $[\lambda]_n=[\lambda]$.

\section{Construction of the compression algorithm}

The state $\ket{\phi}^{\otimes n}$ is symmetric under the action of $\mathbf{P}(s)$ for every $s\in S_n$. By Schur--Weyl duality, it therefore has the form
\begin{align}
\ket{\phi}^{\otimes n}
=\sum_{[\lambda]\in\mathcal{I}_{d,n}}\sum_{q_{[\lambda]},p_{[\lambda]}}
C^{[\lambda]}_{q_{[\lambda]},p_{[\lambda]}}
\ket{[\lambda]}\ket{q_{[\lambda]}}\ket{p_{[\lambda]}}
\end{align}
with support only on permutation-basis vectors satisfying
\begin{align}
\mathbf{p}_{[\lambda]}(s)\ket{p_{[\lambda]}}=\ket{p_{[\lambda]}}
\end{align}
for every $s\in S_n$. The only irreducible representation of $S_n$ containing such an invariant vector is the trivial representation indexed by $[n]$ \cite{james1981representation}. Hence compression of identically prepared states reduces to the Schur transform on the fully symmetric sector. We implement this restricted transform recursively with Clebsch--Gordan (CG) transforms.

Let $\mathcal{Q}_{[\mu]}^d$ and $\mathcal{Q}_{[\nu]}^d$, with $[\mu]\in\mathcal{I}_{d,n}$ and $[\nu]\in\mathcal{I}_{d,m}$, be irreducible representations of $U(d)$. Their tensor product is again a representation of $U(d)$ and decomposes into a direct sum of irreducible representations:
\begin{align}
\mathcal{Q}_{[\mu]}^d \otimes\mathcal{Q}_{[\nu]}^d \stackrel{U(d)}{\cong} \bigoplus_{[\lambda] \in \mathcal{I}_{d,n+m}} \mathcal{Q}_{[\lambda]}^d \otimes \mathcal{M}^{[\lambda]}_{[\mu] ,[\nu]}
\end{align}
where $\mathcal{M}^{[\lambda]}_{[\mu],[\nu]}$ is the multiplicity space associated with $[\mu]$, $[\nu]$, and $[\lambda]$. The transform implementing this isomorphism is the \textit{Clebsch--Gordan transform}.

For compression, we specialize to $[\nu]\equiv[1]$, for which the CG transform is simpler:
\begin{align}
\mathcal{Q}_{[\mu]}^d \otimes\mathcal{Q}_{[1]}^d \stackrel{U(d)}{\cong} \bigoplus_{[\lambda] \in \mu + \square} \mathcal{Q}_{[\lambda]}^d
\end{align} where $[\mu]+\square$ denotes the set of partitions $[\lambda]$ satisfying $\sum_i\lambda_i=n+1$ and $\lambda_i-1\leq\mu_i\leq\lambda_i$. We write $[\lambda]=[\mu]+e_j$ when $\lambda_j=\mu_j+1$, where $e_j$ is the $j$th standard unit vector.

Explicitly, a CG transform $\mathbf{U}_{CG}^d$ acts on a Gelfand--Tsetlin basis state as
\begin{align}
\mathbf{U}^d_{CG}\ket{(\mu)}\ket{i}
=\sum_{[\mu]+e_j\in[\mu]+\square}
\sum_{(\lambda):\,[\lambda]_d=[\mu]+e_j}
C_{(\mu),i}^{(\lambda)}\ket{(\lambda)},
\label{cg}
\end{align}
where $(\lambda)$ ranges over Gelfand--Tsetlin patterns with top row $[\mu]+e_j$, and the coefficients $C_{(\mu),i}^{(\lambda)}\in\mathbb{C}$ depend on the full input and output patterns.

Applying these CG transforms successively constructs the Schur transform. For a computational-basis input $\ket{i_1,\ldots,i_n}$, it suffices to decompose $(\mathcal{Q}_{[1]}^d)^{\otimes n}$ into irreducible representations of $U(d)$; Schur--Weyl duality identifies the corresponding multiplicity spaces with irreducible representations of $S_n$. We perform the decomposition using $n-1$ successive CG transforms. The first combines $\ket{i_1}\in\mathcal{Q}_{[1]}^d$ and $\ket{i_2}\in\mathcal{Q}_{[1]}^d$ to obtain a superposition of $\ket{[1]}\ket{(\lambda)_2}$ as in Eq.~\eqref{cg}. The next CG transform acts on $\ket{(\lambda)_2}\ket{i_3}$ while retaining $\ket{[\lambda]_1}=\ket{[1]}$ as part of the output, producing a superposition of $\ket{[\lambda]_2}\ket{(\lambda)_3}$. Repeating this procedure yields a superposition of partition sequences $\ket{[\lambda]_k}$ together with $\ket{(\lambda)_n}$.

Each output component is labeled by the partition sequence $[\lambda]_1,\ldots,[\lambda]_{n-1}$ together with the final Gelfand--Tsetlin pattern $(\lambda)_n$. Comparing these labels with the Schur basis $\ket{{[\lambda]}}\ket{q_{[\lambda]}}\ket{p_{[\lambda]}}$ gives $\ket{[\lambda]}=\ket{[\lambda]_n}$ and $\ket{q_{[\lambda]}}=\ket{(\lambda)_n}$. Moreover, $[\lambda]_{k+1}=[\lambda]_k+e_{j_k}$ for some $j_k\in[d]$. Hence the full partition sequence specifies a basis vector of the $S_n$ irreducible representation $\mathcal{P}_{[\lambda]}$.

For $d=1$, the CG transform is the trivial map $[m]\mapsto[m+1]$ for $m\in\mathbb{N}_0$. For $d=2$, it maps a tensor-product basis of subsystem states to a basis of eigenvectors of the composite spin operators, and its coefficients can be computed efficiently by a classical algorithm \cite{kirby2017practical}. For general $d$, we use the recursive Clebsch--Gordan construction of Ref.~\cite{bacon2007quantum}, based on the reduced Wigner coefficients of Ref.~\cite{biedenharn1968pattern}, in which $\mathbf{U}^d_{CG}$ is expressed in terms of $\mathbf{U}^{d-1}_{CG}$ and a \textit{reduced Wigner operator}.

The recursive CG transform factors into a $(d-1)$-dimensional CG step followed by a reduced Wigner transform,
\begin{align}
\mathbf{U}^d_{CG}=\Phi\circ\widetilde{\mathbf{U}}^{d-1}_{CG},
\end{align}
where $\widetilde{\mathbf{U}}^{d-1}_{CG}$ acts on the $U(d-1)$ part of the Gelfand--Tsetlin pattern. For $i<d$, this step produces a branch label $j^\prime\in\{1,\ldots,d-1\}$; for $i=d$, the lower-dimensional CG transform is bypassed and we set $j^\prime=0$. Thus $j^\prime\in\{0,\ldots,d-1\}$ in the reduced-Wigner step.

Let $\mu=[\mu]_d$ and $\mu^\prime=[\mu]_{d-1}$ denote the top two rows before the lower-dimensional CG step, and define
\begin{align}
\widetilde{\mu}_j&=\mu_j+d-j, && j\in[d],\\
\widetilde{\mu}^{\prime}_j&=\mu^{\prime}_j+d-1-j, && j\in[d-1].
\end{align}
Translating the reduced-Wigner formula of Biedenharn and Louck \cite{biedenharn1968pattern} into the indexing and phase convention used here gives
\begin{widetext}
\begin{equation}
\begin{split}
\widehat{T}^{\mu,j,\mu^\prime,j^\prime}
=\begin{cases}
S(j^\prime-j)\left[
\dfrac{
\prod\limits_{s\in[d-1]\setminus\{j^\prime\}}(\widetilde{\mu}_j-\widetilde{\mu}^{\prime}_s)
\prod\limits_{t\in[d]\setminus\{j\}}(\widetilde{\mu}^{\prime}_{j^\prime}-\widetilde{\mu}_t+1)
}{
\prod\limits_{s\in[d]\setminus\{j\}}(\widetilde{\mu}_j-\widetilde{\mu}_s)
\prod\limits_{t\in[d-1]\setminus\{j^\prime\}}(\widetilde{\mu}^{\prime}_{j^\prime}-\widetilde{\mu}^{\prime}_t+1)
}
\right]^{1/2},
& j^\prime\in\{1,\ldots,d-1\},\\[4pt]
\left[
\dfrac{
\prod\limits_{s\in[d-1]}(\widetilde{\mu}_j-\widetilde{\mu}^{\prime}_s)
}{
\prod\limits_{s\in[d]\setminus\{j\}}(\widetilde{\mu}_j-\widetilde{\mu}_s)
}
\right]^{1/2},
& j^\prime=0.
\end{cases}
\end{split}
\label{rw}
\end{equation}
\end{widetext}
Here $S(x)=1$ for $x\geq0$ and $S(x)=-1$ for $x<0$. The coefficient $\widehat{T}^{\mu,j,\mu^\prime,j^\prime}$ uses the $U(d-1)$ row $\mu^\prime$ before the lower-dimensional step. A square reduced-Wigner block is instead specified by the top row $\mu$ and the post-step row $\nu$: its column $j^\prime\geq1$ uses $\mu^\prime=\nu-e_{j^\prime}$, while the column $j^\prime=0$ uses $\mu^\prime=\nu$. With this identification, the admissible columns and rows form the unitary block implemented by $\Phi$.

The resulting compression procedure is:
\vspace{9pt}

\noindent{\bf Algorithm 1: Compression} \\
{\bf Inputs:}\\(1) Classical registers storing $d$ and $n$. (2) $n$ identically prepared qudits in the state $\ket{\phi}^{\otimes n}$. \\
{\bf Outputs:} \\A state $\sum_{i=1}^r x_i \ket{i}$, with $r=\dim(\mathcal{Q}_{[n]}^{d})$.\\
{\bf Procedure:}\\
{\bf 1.} For $k=1,...,n-1$:\\
{\bf 2.}\quad Apply one Clebsch--Gordan step to the current $U(d)$-irrep register and the $(k+1)$st qudit.\\
{\bf 3.} End.
\vspace{9pt}

Algorithm 1 uses Schur--Weyl duality to compress identically prepared states. Because the state lies in the symmetric sector, the irrep sequence is fixed to $[\lambda]_k=[k]$ at each recursive step, and $\mathcal{P}_{[n]}$ is trivial. The registers $\ket{[\lambda]}$ and $\ket{p_{[\lambda]}}$ can therefore be omitted. A fixed ordering identifies a basis of $\mathcal{Q}_{[n]}^{d}$ with the first $r$ basis states of the output space.

Let $\mathcal{E}:\operatorname{Sym}^n(\mathbb{C}^d)\to\mathbb{C}^r$ denote this restricted Schur transform followed by the fixed basis relabeling. The decoder $\mathcal{D}$ embeds the $r$-dimensional memory back into $\mathcal{Q}_{[n]}^d$, appends the fixed label $\ket{[n]}$ and the unique trivial $\mathcal{P}_{[n]}$ basis state, and applies the inverse Schur transform. Hence
\begin{align}
\mathcal{D}\mathcal{E}\ket{\phi}^{\otimes n}=\ket{\phi}^{\otimes n}
\label{exactdecode}
\end{align}
for every $\ket{\phi}\in\mathbb{C}^d$. Thus the mathematical encoder and decoder are exact on the input family. Under the fixed identification $\mathbb{C}^r\cong\mathcal{Q}_{[n]}^d$, the encoder also intertwines the collective $U(d)$ action,
\begin{align}
\mathcal{E}U^{\otimes n}=\mathbf{q}_{[n]}^d(U)\mathcal{E},
\label{equivariance}
\end{align}
while permutations act trivially on its symmetric domain.

The CG transform used in Algorithm 1 is implemented as follows.
\vspace{9pt}

\noindent{\bf Algorithm 2: Clebsch--Gordan transform} \\
{\bf Inputs:}\\(1) A classical register storing $d$. (2) A state in $\mathcal{Q}_{[\mu]}^d \otimes\mathcal{Q}_{[1]}^d$, where $[\mu]\in\mathcal{I}_{d,n}$.\\
{\bf Outputs:} \\A state belonging to $\bigoplus_{[\lambda] \in \mu + \square} \mathcal{Q}_{[\lambda]}^d $.\\
{\bf Procedure:}\\
{\bf 1.} For a basis state $\ket{(\mu)}\ket{i}\in\mathcal{Q}_{[\mu]}^d\otimes\mathcal{Q}_{[1]}^d$, if $d=1$:\\
{\bf 2.}\quad For input label $[m]$, output the updated label $[m+1]$.\\
{\bf 3.} Else:\\
{\bf 4.}\quad If $i<d$:\\
{\bf 5.}\quad \quad Perform the $(d-1)$-dimensional CG transform on $\ket{(\mu)_{d-1}} \ket{i}$ and output the superposition of $\ket{(\lambda)_{d-1}} = \ket{\begin{bmatrix}
[\mu]_{d-1}+e_{j^\prime}  \\
(\mu)^{\prime}_{d-2}
\end{bmatrix}}$. Record $j^\prime$.\\
{\bf 6.}\quad Else, if $i=d$:\\
{\bf 7.}\quad \quad Output $\ket{(\lambda)_{d-1}} = \ket{(\mu)_{d-1}}$ and record $j^\prime = 0$.\\
{\bf 8.}\quad Perform the operator $\Phi$ using the reduced Wigner coefficients in Eq.~\eqref{rw}.\\
{\bf 9.} End.

\vspace{9pt}

\section{Optimality and efficient implementation}

Let ${\cal C}_{d,n}=\{(c_1,\dots,c_d):c_1+\cdots+c_d=n,\ c_i\geq0\}$, and let ${\cal O}(c)$ be the set of computational-basis strings containing $c_i$ copies of $\ket{i-1}$. Define the normalized occupation basis
\begin{equation}
\ket{c}=\sqrt{\frac{\prod_{i=1}^d c_i!}{n!}}\sum_{\mathbf{i}\in{\cal O}(c)}\ket{\mathbf{i}}.
\label{occupationbasis}
\end{equation}
For
\begin{equation}
\ket{\phi}=\alpha_1\ket{0}+\cdots+\alpha_d\ket{d-1},
\end{equation}
direct expansion gives
\begin{equation}
\ket{\phi}^{\otimes n}
=\sum_{c\in{\cal C}_{d,n}}
\sqrt{\frac{n!}{\prod_i c_i!}}
\left(\prod_{i=1}^{d}\alpha_i^{c_i}\right)\ket{c}.
\label{occupationexpansion}
\end{equation}
The occupation vectors form an orthonormal basis of $\operatorname{Sym}^n(\mathbb{C}^d)$, and their number is
\begin{equation}
    r=\dim(\mathcal{Q}_{[n]}^d)=\dim\operatorname{Sym}^n(\mathbb{C}^d)=\binom{n+d-1}{d-1}.
\label{memorydim}
\end{equation}
The tensor powers span this symmetric subspace. Indeed, the coefficient of each occupation vector in Eq.~\eqref{occupationexpansion} is a nonzero constant times the homogeneous monomial $\alpha_1^{c_1}\cdots\alpha_d^{c_d}$. If a linear functional vanished on every $\ket{\phi}^{\otimes n}$, the resulting homogeneous polynomial would vanish identically, so all its occupation-basis coefficients would be zero.

This spanning property gives the matching lower bound. Consider any reversible coherent compression with linear encoder $E$ and decoder $D$ satisfying $DE\ket{\phi}^{\otimes n}=\ket{\phi}^{\otimes n}$ for every $\ket{\phi}$. By linearity, $DE$ is the identity on $\operatorname{Sym}^n(\mathbb{C}^d)$, so $E$ is injective on an $r$-dimensional space. Its output Hilbert space must therefore have dimension at least $r$. Algorithm~1 reaches this lower bound and is space-optimal for reversible coherent compression.

The corresponding quantum memory size is
\begin{equation}
    m(n,d)=\left\lceil\log_2\binom{n+d-1}{d-1}\right\rceil
\label{memoryqubits}
\end{equation}
qubits. For fixed $d$,
\begin{align}
    m(n,d)=(d-1)\log_2 n+O(1),
\end{align}
so the number of qubits grows only logarithmically with $n$.

The symmetric Clebsch--Gordan branch has a simple form in the occupation basis. For $c\in{\cal C}_{d,k}$, let $\Pi_{k+1}$ be the projector from $\operatorname{Sym}^k(\mathbb{C}^d)\otimes\mathbb{C}^d$ onto $\operatorname{Sym}^{k+1}(\mathbb{C}^d)$. Then
\begin{equation}
\Pi_{k+1}\bigl(\ket{c}\ket{i}\bigr)
=\sqrt{\frac{c_i+1}{k+1}}\,\ket{c+e_i}.
\label{symmetriccg}
\end{equation}
Indeed, all strings in $\ket{c+e_i}$ have equal amplitude, and the fraction whose last symbol is $i$ is $(c_i+1)/(k+1)$. This gives the overlap in Eq.~\eqref{symmetriccg}. The remaining component belongs to the orthogonal Clebsch--Gordan branches.

At the $k$th compression step, identify $\mathcal{Q}_{[k]}^d$ with $\operatorname{Sym}^k(\mathbb{C}^d)$ and define
\begin{equation}
\mathcal{R}_{k,d}=\operatorname{span}\left\{\ket{\phi}^{\otimes k}\otimes\ket{\phi}:\ket{\phi}\in\mathbb{C}^d\right\}
\subset \mathcal{Q}_{[k]}^d\otimes\mathcal{Q}_{[1]}^d.
\label{relevantsubspace}
\end{equation}
The full Clebsch--Gordan transform sends $\mathcal{R}_{k,d}$ entirely into $\mathcal{Q}_{[k+1]}^d$. For a one-row symmetric representation, the $U(r)$ row of the Gelfand--Tsetlin pattern records the cumulative occupation $[c_1+\cdots+c_r]$. Let $[\ell]$ be the post-step $U(d-1)$ row. The two relevant columns use the pre-step rows $[\ell-1]$ for $j^\prime=1$ and $[\ell]$ for $j^\prime=0$. Substituting these one-row partitions into Eq.~\eqref{rw} gives the two coefficients for the symmetric output $j=1$:
\begin{equation}
 a_\ell=\sqrt{\frac{\ell}{k+1}},\qquad
 b_\ell=\sqrt{\frac{k+1-\ell}{k+1}},
\label{twocoefficients}
\end{equation}
corresponding respectively to $j^\prime=1$ and $j^\prime=0$. Since $a_\ell^2+b_\ell^2=1$, they form the first row of the two-level rotation
\begin{equation}
G_\ell=
\begin{pmatrix}
 a_\ell & b_\ell\\
 -b_\ell & a_\ell
\end{pmatrix}.
\label{givens}
\end{equation}
Thus the restricted recursion needs only a controlled Givens rotation on the two branch labels reached by symmetric inputs.

\vspace{9pt}
\noindent{\bf Algorithm 2': Restricted Clebsch--Gordan step} \\
{\bf Inputs:}\\(1) Classical registers storing $d$ and $k$. (2) A state promised to lie in $\mathcal{R}_{k,d}$.\\
{\bf Outputs:} \\The corresponding state in $\mathcal{Q}_{[k+1]}^d$.\\
{\bf Procedure:}\\
{\bf 1.} If $d=1$, apply the trivial one-dimensional Clebsch--Gordan map and stop.\\
{\bf 2.} Unpack the Gelfand--Tsetlin register into its $U(d-1)$ row and lower-dimensional pattern.\\
{\bf 3.} Coherently, if $i<d$, apply Algorithm~2' in dimension $d-1$ to the lower-dimensional symmetric component and record $j^\prime=1$. If $i=d$, leave that component unchanged and record $j^\prime=0$.\\
{\bf 4.} Let $[\ell]$ be the resulting $U(d-1)$ row and compute $a_\ell,b_\ell$ from Eq.~\eqref{twocoefficients}.\\
{\bf 5.} Apply the controlled two-level rotation $G_\ell$ of Eq.~\eqref{givens} to the branch labels $j^\prime=1,0$, extending it by the identity on unused labels. On the promised input, retain the symmetric output label $j=1$.\\
{\bf 6.} Repack the Gelfand--Tsetlin register.\\
{\bf 7.} End.\\
\vspace{9pt}

\begin{proposition}
For every $k\geq1$ and $d\geq1$, Algorithm~2' agrees with the full Clebsch--Gordan transform on $\mathcal{R}_{k,d}$. Consequently, the successive restricted steps in Algorithm~1 produce exactly the same compressed state as the full Schur transform on every input $\ket{\phi}^{\otimes n}$.
\label{prop:restrictedcg}
\end{proposition}

\emph{Proof.}
We use induction on $d$. The case $d=1$ is immediate. Assume the claim in dimension $d-1$ and consider an occupation vector $c=(c_1,\ldots,c_d)$ with $\sum_i c_i=k$.

If the new local label satisfies $i<d$, the lower-dimensional step acts on the first $d-1$ levels. Let $\ell=\sum_{a=1}^{d-1}c_a+1$ be their occupation after adding the new symbol. By the induction hypothesis, the symmetric coefficient from that step is $\sqrt{(c_i+1)/\ell}$. The top-level rotation contributes $a_\ell=\sqrt{\ell/(k+1)}$, so their product is $\sqrt{(c_i+1)/(k+1)}$. If $i=d$, the lower-dimensional step is bypassed, $\ell=\sum_{a=1}^{d-1}c_a$, and the top-level coefficient is
\begin{equation}
 b_\ell=\sqrt{\frac{k+1-\ell}{k+1}}
 =\sqrt{\frac{c_d+1}{k+1}}.
\end{equation}
Thus Algorithm~2' has exactly the symmetric projection in Eq.~\eqref{symmetriccg} for every occupation-basis component.

For any $x\in\mathcal{R}_{k,d}$, the full Clebsch--Gordan transform has no component outside the $[k+1]$ sector. Algorithm~2' is unitary and has the same $[k+1]$ component; that component already has norm $\|x\|$, so no orthogonal component remains. Hence the two transforms agree on $\mathcal{R}_{k,d}$. Cascading the restricted steps gives the stated Schur transform on $\ket{\phi}^{\otimes n}$. \hfill$\square$

Proposition~\ref{prop:restrictedcg} is exact at the mathematical level. Over a fixed finite gate set, the required rotations can instead be approximated to any operator-norm accuracy. The standard recursive Clebsch--Gordan construction computes these rotations efficiently, with a per-step upper bound $d^3\operatorname{poly}(\log n,\log(1/\delta))$ for accuracy $\delta$ \cite{bacon2007quantum}. Since Algorithm~2' uses only the controlled rotations in Eq.~\eqref{givens}, the same upper bound applies.

\begin{proposition}
For every $\epsilon\in(0,1)$, Algorithm~1 can be implemented over a fixed finite universal gate set with operator-norm error at most $\epsilon$ using
\begin{equation}
 n d^3\,\operatorname{poly}\!\left(\log n,\log\frac{n}{\epsilon}\right)
\label{complexity}
\end{equation}
elementary one- and two-qubit gates. The decoder has the same asymptotic complexity. In particular, for fixed $d$ the encoder and decoder have polynomial complexity in $n$ and $\log(1/\epsilon)$.
\label{prop:complexity}
\end{proposition}

\emph{Proof.}
There are $n-1$ recursive Clebsch--Gordan steps. Approximate each step within operator norm $\delta=\epsilon/(n-1)$. The standard reduced-Wigner construction gives a per-step cost $d^3\operatorname{poly}(\log n,\log(1/\delta))$ \cite{bacon2007quantum}. For products of unitaries, a telescoping expansion bounds the total operator-norm error by the sum of the individual errors, so the encoder error is at most $(n-1)\delta=\epsilon$. Substituting $\delta$ gives Eq.~\eqref{complexity}. The decoder applies the inverse sequence and has the same gate count. \hfill$\square$

\section{Conclusion}
The key structural fact is that $n$ identical copies of a pure $d$-dimensional state occupy the fully symmetric sector of the Schur--Weyl decomposition. Restricting the Schur transform to this sector removes the fixed Schur label and the trivial permutation register while retaining exactly the information needed for recovery. The symmetric subspace has dimension $\binom{n+d-1}{d-1}$, and tensor-power states span it, so this dimension is both achievable and necessary for reversible coherent compression. The inverse restricted transform gives exact decoding.

At each Clebsch--Gordan step, the recursive implementation only follows the branch leading to the next symmetric representation. In the occupation basis this branch is realized by a controlled two-level rotation, and the restricted recursion agrees with the full Schur transform on the admissible input family. Over a fixed finite universal gate set, standard Clebsch--Gordan synthesis approximates these rotations to arbitrary accuracy with polynomial complexity. Thus the construction gives lossless, space-optimal compression together with an efficient implementation for arbitrary fixed local dimension.

\section{Acknowledgments} This work is supported by the National Natural Science Foundation of China under Grant Nos. 61877054, 12031004, 12271474, and 12201555, and by the Foreign Experts in Culture and Education Foundation under Grant No. DL2022147005L. Kamil Khadiev acknowledges support from these projects for his visit.

\end{document}